\documentstyle[12pt,epsfig]{article}
\begin{document}

\newcommand{\al}{\mbox{$\alpha$}}
\newcommand{\be}{\mbox{$\beta}}
\newcommand{\g}{\mbox{$\gamma$}}
\newcommand{\de}{\mbox{$\delta$}}
\newcommand{\om}{\mbox{$\omega$}}
\newcommand{\Om}{\mbox{$\Omega$}}
\newcommand{\tri}{\mbox{$\triangle$}}
\newcommand{\tr}{\mbox{$\tilde{r}$}}
\newcommand{\tom}{\mbox{$\tilde{\omega }$}}

\def\beq{\begin{equation}}
\def\eeq{\end{equation}}
\def\beqa{\begin{eqnarray}}
\def\eeqa{\end{eqnarray}}
\def\s{\sigma}   
\def\ve{\varepsilon}
\def\bl{\bar{\lambda}}
\def\da{\dagger}
\def\la{\lambda}   
\def\ti{\tilde}
\def\pr{\prime} 
\def\f{\frac}
\def\sq{\sqrt}
\def\pa{\partial}

\begin{flushright} 
gr-qc/9705015  \\
\end{flushright}
\begin{center}
   \vskip 3em 
   {\LARGE Quantum Ergoregion Instability\footnote{To appear in the 
    Proceedings of the APCTP Winter School on Duality of String Theory,
    Korea, Feb. 17-28, 1997.}}
   \vskip 1.5em
   {\large Gungwon Kang   
   \\[.5em]}
{\em Raman Research Institute, Bangalore 560 080, 
India; kang@rri.ernet.in}  \\[.7em]
\end{center}
\vskip 1em

\noindent 
{\bf Abstract: } We have shown that, as in the case of black holes, 
an ergosphere itself with no event horizon inside can evaporate spontaneously,
giving energy radiation to spatial infinity until the ergoregion disappears.
However, the feature of this quantum ergoregion instability is very much 
different from black hole radiation. It is rather analogous to a laser 
amplification. This analysis is based on the canonical quantization of a 
neutral scalar field in the presence of unstable modes characterized by 
complex frequencies in a simple model for a rapidly rotating star. 

\vskip .5cm
In general relativity, inertial frames around a rotating object are dragged 
in the sense of the rotation. If the object is rotating rapidly, this 
dragging effect can be so strong that in some region no physical object 
can remain at rest relative to an inertial observer at spatial infinity. 
This region of spacetime is called an ergoregion or an ergosphere. 
The most common example of ergoregions would be the outside of the 
event horizon of any rotating black hole. Ergoregions also arise in models 
of dense, rotating stars, in which cases no event horizon exists inside the 
ergosurface \cite{BI,SC}. The appearance of any ergoregion causes the 
well-known classical amplification phenomena such as Penrose process 
for particles and
superradiance for waves. Let us consider a regular matter distribution 
with non-zero total angular momentum in the past such that the spacetime 
is almost flat. As the gravitational collapse occurs, the system may end 
up into a stationary rotating black hole in the future. As is well-known,
quantum field theory in this evolving background spacetime shows that the
black hole formed radiates spontaneously to infinity. In other words, there
is no stable vacuum for matter fields in the presence of a rotating black 
hole which contains both event horizon and ergosphere. 
On the other hand, suppose the collapsing matter ends up into a stationary 
rotating star, which has an {\it ergoregion but no horizon} in the future. 
Now one may ask whether quantum field theory in this background spacetime 
also gives rise to a spontaneous radiation of the ergosphere as in the 
case of a collapsing black hole. As argued in Ref.~\cite{Ford}, 
an ergosphere may not evaporate spontaneously since the transmitted part
of an ingoing spherical wave packet in a superradiant mode, which carries 
a negative energy with respect to an observer at infinity, 
will come out again if there is no event horizon inside 
the ergosurface after passing the center of a rotating object, leaving no 
net gain of positive energy at infinity. 
Recently, Matacz, Davies, and Ottewill \cite{MDO} have considered 
a quantum scalar field in a simple model describing such spacetimes.
One of the motivations of their study is to isolate the role of the 
ergoregion in order to resolve whether 
the Starobinskii-Unruh effect \cite{Staro,Unruh,Ford,AM1} is primarily
due to the existence of an event horizon or an ergoregion. They showed
that the quantum vacuum for a scalar field is {\it stable} if all mode 
frequencies are real, indicating no Starobinskii-Unruh effect in the
presence of ergoregion only. This result, however, contradicts with 
some general results of Ashtekar and Magnon \cite{AM2,AM1}
obtained in the algebraic approach to quantization. 
In addition, spacetimes with ergoregions but no horizons such as rapidly 
rotating stars are known to be unstable to classical scalar, 
electromagnetic, and gravitational perturbations \cite{Fried,CS,Vilen}. 
Such classical instability, the so-called ergoregion instability, can be 
intuitively explained as follows. After passing through the center of the 
rotating object, the transmitted part, carrying negative energy, of an 
incident wave packet in a ``superradiant" mode will scatter at the 
ergosurface, giving transmission
as well as reflection there with energies amplified, respectively. 
This process will repeat as long as the ergoregion remains, resulting in 
presumably  ``exponential" radiation of positive energy to infinity 
and accumulation of negative energy within the ergoregion in such a way that
the total energy is conserved. It turns out that this instability is
characterized by complex frequency modes in normal mode solutions of
classical fields. Therefore, not all mode frequencies are real in such a
background spacetime. 

In this paper, we incorporate such unstable complex frequency modes
into the field quantization and show that they lead to a spontaneous 
evaporation of an ergoregion. This quantum ergoregion instability is 
very much analogous to a laser amplification. The extended version of
this article can be found in Ref.~\cite{Kang2}. 

\vskip .5cm 
Now let us consider a system of a massless real scalar field 
$\phi (x)$ minimally 
coupled to gravitational fields satisfying the Klein-Gordon equations
given by
\beq
\Box \phi = \frac{1}{\sq{-g}}\pa_{\mu}(\sq{-g}g^{\mu \nu}
\pa_{\nu}\phi )=0. 
\label{KG}
\eeq
We assume that the background spacetime possesses a Killing vector 
field $\xi =\pa_t$ in some regions, for instance, in the early and 
late stages of its evolution. 
Then normal mode solutions can be defined by
${\cal L}_\xi \phi = -i \om \phi $. 
Here ${\cal L}_\xi$ is the Lie-derivative along a Killing vector $\xi$.
Since ${\cal L}_\xi \phi = \pa \phi /{\pa x^0}$,
the time dependence of normal mode solutions is $\sim e^{-i\om t}$.
Given the usual Klein-Gordon inner product  
\beq
\langle \! \phi_1\, ,\, \phi_2\! \rangle
 = \frac{i}{2}\int \, \phi_1^{\ast}\! \stackrel{\leftrightarrow }
     {\partial_{\mu}}\! \phi_2 \,\, d\Sigma^{\mu} 
\label{Inner}
\eeq
at an $x^0=t={\rm const}$ spacelike hypersurface, one obtains 
\beq
(\om_2-\om_1^{\ast})\langle \! \phi_1\, ,\, \phi_2\! \rangle = 0
\label{OrthoG}
\eeq
for any given two normal mode solutions $\phi_1$ and $\phi_2$: 
see Ref.~\cite{Kang2} for details. Thus the inner
product is zero unless $\om_2=\om_1^{\ast}$. Since our inner product defined 
in  Eq.~(\ref{Inner}) is not positive definite in general, the normal
mode frequency $\om$ is not necessarily always real. 
It is indeed possible that there exist bounded solutions with
complex frequencies as shown in Refs. \cite{Vilen,CS,Fried} for certain 
cases of spacetimes with ergoregions. 
Then, from Eq.~(\ref{OrthoG}), the norm of such complex 
frequency modes should be zero. 

We assume that our background spacetime is described by the Minkowski 
flat metric in the past infinity and by the Kerr metric with mirror 
boundary condition on the field $\phi$, which is used in 
Refs.~\cite{MDO,Vilen}, in the future infinity: Instead of considering 
the detailed dynamics inside the star, we simply assume that all classical
solutions $\phi$ of Eq.~(\ref{KG}) vanish on the surface of 
some sphere inside the 
ergoregion, e.g., a totally reflecting mirror boundary. The quantization 
of the field in the past infinity will be straightforward; it will have 
a Fock representation with a vacuum state $|0\! \rangle_{\!\! \rm in}$. 
To carry out the canonical quantization in the future infinity, let us 
first construct normal mode solutions of Eq.~(\ref{KG}). 

As is well known, the Klein-Gordon equation is 
separable \cite{SEP} and, in Boyer-Linquist coordinates, admits 
a complete set of normal mode solutions of the form 
\beq
\phi (x) =\f{R(r)}{\sq{r^2+a^2}}S(\theta )e^{-i\om t+im\varphi }, 
\eeq
where $a$ is the angular momentum per unit mass of the star with mass
$M$, and $m$ is an integer. 
Defining a ``tortoise" coordinate $\ti{r}$ by
$d\ti{r}/{dr}=(r^2+a^2)/(r^2+a^2-2Mr)$, 
the radial part of Eq.~(\ref{KG}) becomes 
\beq
\f{d^2R}{d\ti{r}^2} - V_{\om lm}(\ti{r})R = 0.
\label{KGR2}
\eeq
From the mirror boundary condition, the radial function vanishes at some 
$r=r_0$ (accordingly, $\ti{r}=\ti{r}_0$) inside the ergoregion:  
$R(r_0)=0$.
For simplicity, we assume that $r_0$ is very near the ``horizon" radius
$r=r_{\rm H}=M+\sq{M^2-a^2}$. That is, $\ti{r}_0 \sim -\infty $. 
We also require that the field is not singular at spatial infinity, 
$\ti{r} \sim \infty $. 
The asymptotic behavior of the effective potential $V_{\om lm}$
induced through the interaction with gravitational fields is 
\cite{footnote1}
\beq
V_{\omega lm}(r) \sim \left\{\begin{array}{ll}
                               -(\omega -m\Omega_H)^2  & 
\mbox{as $\tilde{r} \rightarrow \tilde{r}_0$,}    \\ 
                           -\omega^2     \qquad   &
\mbox{as $\tilde{r} \rightarrow \infty $,} 
			     \end{array}
                      \right. 
\label{Poten}
\eeq
where $\Om_{\rm H}=a/{2Mr_{\rm H}}$. We have $V_{\om lm} = \infty $ 
at $\ti{r} = \ti{r}_0$, corresponding to the mirror boundary condition. 
Since $\om$ could be complex in our model,
$V_{\omega lm}(r)$ is a complex potential in general.

Let us now consider normal mode solutions $u_{\om lm}(r)$ to Eq.~(\ref{KGR2})
whose asymptotic forms are
\beq
u_{\omega lm}(r) \sim \left\{\begin{array}{ll}
                 B_{\om lm}(e^{i\ti{\om}\ti{r}} 
                 +A_{\om lm}e^{-i\ti{\om}\ti{r}}) & 
\mbox{as $\ti{r} \rightarrow \ti{r}_0$,}    \\ 
                 e^{-i\om \ti{r}} + C_{\om lm}e^{i\om \ti{r}}  &               
\mbox{as $\ti{r} \rightarrow \infty $,} 
			     \end{array}
                      \right.
\label{ASolR}
\eeq
where $A_{\om lm}=-e^{2i\tom \tr_0}$ 
from the mirror boundary condition at $\tr =\tr_0$.  
If $\om$ is complex, $u_{\om lm}$ becomes exponentially divergent at spatial 
infinity and so we exclude this class of solutions from our construction. 
Thus $u_{\om lm}$ represents real frequency normal mode solutions. 
Now the Wronskian relations from  Eq.~(\ref{KGR2}) with the mirror boundary 
condition give
$|A_{\om lm}| = |C_{\om lm}| = 1$. 
Therefore, $u_{\om lm}(r)$ is a stationary wave without any net ingoing or 
outgoing flux with respect to ZAMO's \cite{ZAMO}. Here $\om$
is any continuous real number. In fact, this class of real frequency normal 
mode solutions is equivalent to the set considered in Ref.~\cite{MDO} 
as a complete basis. 

As mentioned above, however, there exists another class of normal mode 
solutions with complex frequencies which describe unstable modes in the 
presence of ergoregions. This class of solutions has not been included 
in the quantization procedure in Ref.~\cite{MDO}. Let $v_{\om lm}(r)$ be 
normal mode solutions in such class whose asymptotic behaviors are   
\beq
v_{\om lm}(r) \sim \left\{\begin{array}{ll}
                 e^{i\ti{\om}\ti{r}} +R_{\om lm}e^{-i\tom \tr}  & 
\mbox{as $\ti{r} \rightarrow \ti{r}_0$,}    \\ 
                 T_{\om lm}e^{i\om \ti{r}}   \qquad   &
\mbox{as $\ti{r} \rightarrow \infty $.} 
			     \end{array}
                      \right. 
\label{ASolC}
\eeq
The mirror boundary condition is satisfied if
\beq
R_{\om lm}=-e^{2i\tom \tr_0}=-e^{2i\tom^{\rm R}\tr_0}\cdot 
	    e^{-2\tom^{\rm I}\tr_0},
\eeq
where $\tom =\tom^{\rm R}+i\tom^{\rm I}$. Thus, 
$\om^{\rm I}=\tom^{\rm I}=-\ln |R_{\om lm}|/{2\tr_0}$.
Note that, since the potential 
$V_{\om lm}(r)$ in Eq.~(\ref{KGR2}) is complex, the Wronskian relation
does not necessarily give $|R_{\om lm}|=1$. 
If $|R_{\om lm}| > 1$, $\om^{\rm I} > 0$ and so $v_{\om lm}(r) \sim 
e^{-\om^{\rm I}\tr}$ as $\tr \rightarrow \infty$ and is
regular at spatial infinity. From the time dependence
of this solution, i.e., $\sim e^{-i\om t}=e^{-i\om^{\rm R}t}\cdot 
e^{\om^{\rm I}t}$, we also notice that it represents an outgoing mode 
which is exponentially amplifying in time but is exponentially 
decreasing as $\tr \rightarrow \infty$. By making a wave packet, 
as suggested in Ref.~\cite{Vilen}, we may regard this solution as an 
outgoing wave packet with $\tom^{\rm R}=\om^{\rm R}-m\Om_{\rm H} < 0$
starting from near the mirror surface, which will bounce back and forth
within the ergoregion, and a part of which is repeatedly transmitted
to infinity, resulting in exponential amplification in time in the 
inside as well as on the outside of the ergosurface. If $|R_{\om lm}| < 1$,
$\om^{\rm I} < 0$ and so this solution corresponds to an outgoing 
decaying mode in time. However, since its radial behavior becomes 
singular at spatial infinity, we do not include this mode 
in our construction of normal mode solutions.  

For any given solution
\beq
\phi_{\om lm}(x) =\phi_{\om lm}(r, \theta )e^{-i\om t+im\varphi }
=\f{v_{\om lm}(r)}{\sq{r^2+a^2}}S_{\om lm}(\theta )e^{-i\om t+im\varphi}
\sim e^{\om^{\rm I}t}, 
\eeq
with $\om^{\rm I}>0$, we find that there are three linearly independent 
solutions: 
\beqa
\phi^{\ast}_{\om lm}(x) &=& \f{v^{\ast}_{\om lm}(r)}{\sq{r^2+a^2}}
	S^{\ast}_{\om lm}(\theta )e^{i\om^{\ast}t-im\varphi }
	\sim e^{\om^{\rm I}t},   \nonumber     \\
\phi_{\om^{\ast}lm}(x) &=& \f{v^{\ast}_{\om lm}(r)}{\sq{r^2+a^2}}
	S^{\ast}_{\om lm}(\theta )e^{-i\om^{\ast}t+im\varphi }
	\sim e^{-\om^{\rm I}t},   \nonumber     \\	
\phi^{\ast}_{\om^{\ast}lm}(x) &=& \f{v_{\om lm}(r)}{\sq{r^2+a^2}}
	S_{\om lm}(\theta )e^{i\om t-im\varphi }
	\sim e^{-\om^{\rm I}t}.   
\eeqa
$\phi_{\om^{\ast}lm}$ 
represents an exponentially dacaying wave in time which 
originates at infinity. In other words, this mode is the same as 
$\phi_{\om lm}(x)$ but backward in time. 
For these non-stationary modes, $\om$ is a discrete complex 
number which is determined by the details of the potential and the
boundary condition. In fact, by finding poles of the scattering 
amplitude for a more realistic model of rotating stars, Comins and 
Schutz \cite{CS} have shown that the imaginary part of the complex 
frequency for a purely outgoing mode is discrete, positive, 
and proportional to $e^{-2\beta m}$, where $\beta$ is of order unit. 
For our model, it also can be shown
that complex eigenfrequencies are confined 
to a bounded region. 

From our definition of the inner product in Eq.~(\ref{Inner}), we find
\beq
\langle \! \phi_{\om lm}\, ,\, \phi_{\om^{\pr}l^{\pr}m^{\pr}}\! \rangle 
 = \frac{1}{2}\int \, [(\om^{\pr}+\om^{\ast})-\Om (m'+m)] 
     \phi_{\om lm}^{\ast}\phi_{\om^{\pr}l^{\pr}m^{\pr}} N^{-1}d\Sigma ,
\eeq
where we have used $d\Sigma^{\mu}=N^{-1}(\pa_t+\Om \pa_{\varphi})^{\mu}
d\Sigma$, $\,\, \Om (r, \theta )=-g_{t\varphi}/g_{\varphi \varphi}$, 
and $N=(-g^{tt})^{-1/2}$. 
For real frequency modes 
$u_{\om lm}(x) = u_{\om lm}(r)/\sq{r^2+a^2}S_{\om lm}(\theta )
		e^{-i\om t+im\varphi}$   
with $\om > 0$, we have the orthogonality relations
\beqa
\langle \! u_{\om lm}\, ,\, u_{\om^{\pr}l^{\pr}m^{\pr}}\! \rangle = 
\delta (\om -\om^{\pr})\delta_{ll^{\pr}}\delta_{mm^{\pr}}   \qquad
{\rm for} \qquad  u_{\om lm} \not\in  N^{-},   \nonumber    \\
\langle \! u_{-\om l-m}\, ,\, u_{-\om^{\pr}l^{\pr}-m^{\pr}}\! \rangle = 
\delta (\om -\om^{\pr})\delta_{ll^{\pr}}\delta_{mm^{\pr}}   \qquad
{\rm for} \qquad u_{\om lm} \in  N^{-},
\label{OrthoR}
\eeqa
after suitable normalizations. Here $N^{-}$ is defined as a set 
consisting of mode solutions $u_{\om lm}$ with $\om >0$ whose norms 
are negative.
For complex frequency normal mode solutions
$v_{\om lm}(x) 
=v_{\om lm}(r)/\sq{r^2+a^2}S_{\om lm}(\theta )e^{-i\om t+im\varphi}$,
we have 
\beq
\langle \! v_{\om lm}\, ,\, v_{\om lm}\! \rangle = 0, \,\, 
\langle \! v_{\om lm}\, ,\, v_{\om^{\ast}lm}\! \rangle = 
-\langle \! v^{\ast}_{\om lm}\, ,\, v^{\ast}_{\om^{\ast}lm}\! \rangle =1.
\label{OrthoC}
\eeq
All other inner products vanish. 

Finally, it should be pointed out that it may be possible that 
any normal mode solution with
complex frequency is expressed by linearly combining the real
frequency normal modes $\{ u_{\om lm}(x)\} $. 
There is no disproof for this possibility yet. But, we assume that 
the set of complex frequency normal mode solutions 
represents new independent degrees of freedom of the system, which can
describe field solutions carrying arbitrary values of energy including 
negative ones by linear combinations.

\vskip .5cm 
Based on the analysis of normal mode solutions for the classical scalar 
field above, we now apply the quantization methods developed in 
Refs.~\cite{Schroer,SS,Fulling,Kang} in the presence of complex frequency
modes in the Minkowski flat spacetime.
The neutral scalar field can be expanded in terms of normal mode 
solutions as 
\beqa
\phi (x) &=& \sum_{lm}\int_{\not\in N^{-}} d\om \f{1}{\sq{2}}
	     [a_{\la}u_{\la}(x)+a^{\dagger}_{\la}u^{\ast}_{\la}(x)]
	     +\sum_{lm}\int_{\in N^{-}} d\om \f{1}{\sq{2}}
	     [a_{-\la}u_{-\la}(x)+a^{\dagger}_{-\la}u^{\ast}_{-\la}(x)]
	     \nonumber   \\
	 & & +\sum_{\om lm} \f{1}{\sq{2}}
	     [b_{\la}v_{\la}(x)+b^{\dagger}_{\la}v^{\ast}_{\la}(x)
	      +b_{\bl}v_{\bl}(x)+b^{\da}_{\bl}v^{\ast}_{\bl}(x)],
\label{Field}
\eeqa
where $\la$ denotes $(\om ,l,m)$, $-\la$ denotes $(-\om ,l,-m)$, and 
$\bl$ denotes $(\om^{\ast},l,m)$. 
Assuming the equal-time commutation relations for $\phi (x)$ and 
its momentum conjugate $\pi (x)$, 
we find commutation relations among mode operators
\beq
\lbrack a_{\la}\, ,\, a^{\dagger }_{\la^{\pr}} \rbrack =
\delta_{\la \la^{\pr}}~, \qquad 
\lbrack b_{\la}\, ,\, b^{\dagger }_{\bar{\la^{\pr}}} \rbrack =
\delta_{\la \la^{\pr}}~, \qquad    
\lbrack b_{\la}\, ,\, b^{\dagger }_{\la^{\pr}} \rbrack =
\lbrack b_{\bl}\, ,\, b^{\dagger }_{\bar{\la^{\pr}}} \rbrack =
\lbrack b_{\la}\, ,\, b_{\bar{\la^{\pr}}} \rbrack =0 ~.  
\eeq
All others vanish. Note that the real frequency mode operators satisfy
the usual commutation relations whereas mode operators for complex 
frequencies have unusual commutation relations.
In particular, $b_{\la}$ does commute with $b^{\da}_{\la}$. 

Now the Hamiltonian operator can be expressed in terms of mode 
operators 
\beqa
H &=& \f{1}{2} \sum_{lm}\int_{\not\in N^{-}} d\om \om (a^{\da}_{\la}
      a_{\la}+a_{\la}a^{\da}_{\la}) 
      + \f{1}{2} \sum_{lm}\int_{\in N^{-}} d\om (-\om )(a^{\da}_{-\la}
      a_{-\la}+a_{-\la}a^{\da}_{-\la})   \nonumber    \\
  & & +\f{1}{2}\sum_{\om lm}[\om (b_{\la}b^{\da}_{\bl}+b^{\da}_{\bl}
      b_{\la}) + \om^{\ast} (b^{\da}_{\la}b_{\bl}+b_{\bl}b^{\da}_{\la})],
\label{HamOp1}
\eeqa
where $\om >0$ for real frequency modes and $\om^{\rm I}>0$ for complex
frequency modes. 
The Hamiltonian for real frequency modes has a representation 
of a set of {\it attractive} harmonic oscillators as usual. 
Interestingly, although a vacuum state
can be defined such that $a_{\pm \la}|0\! \rangle_{\!\! \rm R}=0$
for all $\la$, it is not the lowest energy state  
since the energy associated with the second term in Eq.~(\ref{HamOp1}) 
is not bounded below but bounded above. Therefore, real frequency mode 
operators possess the usual symmetrized Fock representation 
${\cal H}^{\rm R}$ as well as the particle interpretation.
For complex frequency modes, let 
\beq
H^C_{\la} =\f{1}{2}[\om (b_{\la}b^{\da}_{\bl}+b^{\da}_{\bl}
      b_{\la}) + \om^{\ast} (b^{\da}_{\la}b_{\bl}+b_{\bl}b^{\da}_{\la})].
\eeq
By using linear transformations into Hermitian operators $q$ and $p$ 
satisfying $\lbrack q, p \rbrack = i$ 
\beqa
b_{\la} &=& \f{1}{2}[i(\sq{\om^{\rm I}} q_{1\la}+\f{1}{\sq{\om^{\rm I}}}
	    p_{1\la}) + (\sq{\om^{\rm I}}q_{2\la}+\f{1}{\sq{\om^{\rm I}}}
	    p_{2\la})],    \nonumber   \\
b^{\da}_{\bl} &=& \f{1}{2}[(\sq{\om^{\rm I}} q_{1\la}-\f{1}{\sq{\om^{\rm I}}}
	    p_{1\la}) + i(\sq{\om^{\rm I}}q_{2\la}-\f{1}{\sq{\om^{\rm I}}}
	    p_{2\la})],
\eeqa
one finds
\beq
H^C_{\la} = \f{1}{2}(p^2_{1\la}-(\om^{\rm I})^2q^2_{1\la})
            +\f{1}{2}(p^2_{2\la}-(\om^{\rm I})^2q^2_{2\la})
	    +\om^{\rm R}(q_{1\la}p_{2\la}-p_{1\la}q_{2\la}).
\eeq
Thus this is a system of two coupled {\it inverted} harmonic oscillators
with the same frequency $|\om |$. 
Consequently, the energy spectrum for $H^C_{\la}$ is
$E_{\ve_{\la}k_{\la}}=\om^{\rm I}\ve_{\la}+\om^{\rm R}k_{\la}$ 
where $\ve_{\la}$ is any continuous real number and $k_{\la}$ any integer.
It shows that
the energy eigenvalue is {\it continuous} for given $k_{\la}$ 
and {\it unbounded} below. This unboundedness of the energy indicates 
that energy eigenstates are not normalizable. However, 
we can construct normalizable wave packets from them. These square
integrable wave packets will form a Hilbert space 
${\cal H}^C_{\la}$ which is isomorphic to $L^2({\bf R}^2)$. 

Any quantum state of the field which is in this Hilbert space will give 
rise to instability. It follows because, although the total energy of this
state is definite and time independent, the energy density outside the 
ergoregion will be positive and have exponential time dependence 
whereas the energy density within the ergoregion will have the same 
behavior but with negative energy, keeping the total energy over
the whole space fixed. Therefore, an observer sitting outside 
the ergoregion will measure time-dependent radiation of positive energy. 
In addition, since the energy spectrum is unbounded below, some external
interaction with this system can give energy extraction from the system
without bound. Note that mode operators no longer have particle 
interpretation as in those for real frequency modes. 
Finally, we complete our quantization of the field $\phi (x)$ 
by constructing the total Hilbert space as follows,
\beq
{\cal H} = {\cal H}^{\rm R} \otimes \prod_{\la} {\cal H}^C_{\la}.
\eeq
Here ${\cal H}^{\rm R}$ is the usual symmetrized Fock space generated
by real frequency modes and $\prod_{\la} {\cal H}^C_{\la}$ is the infinite
number of products of non-Fock-like Hilbert spaces ${\cal H}^C_{\la}$ 
generated by complex frequency modes. 

Now let us see how the appearance of an ergoregion at the late stage of
a dynamically evolving background spacetime starts to give a spontaneous 
radiation of energy. We expect this spontaneous quantum radiation 
if the initial vacuum state $|0\rangle_{\!\! \rm in}$ of the field 
in the past falls in  any state in $\prod_{\la} {\cal H}^C_{\la}$    
in the remote future. To see this effect let us consider 
a ``particle" detector linearly coupled to the field near 
$t \sim \infty$ placed in the in-vacuum state 
$|0\rangle_{\!\! \rm in}$. 
Then, the transition rate of the detector integrated over angular variables
$\theta$ and $\varphi$ is 
proportional to \cite{BD} 
\beqa
\f{F(E)}{T} &\sim & \f{1}{2}\sum_{\s} \{\sum_{lm}
	\int_{\not\in N^{-}} d\om |\beta_{\la \s}|^2
	|\f{u_{\la}(r)}{\sq{r^2+a^2}}|^2\delta (E-\om ) 
	\nonumber   \\
& & + \sum_{lm}\int_{\in N^{-}} d\om
	|\al_{-\la \s}|^2|\f{u_{-\la}(r)}{\sq{r^2+a^2}}|^2
	\delta (E-\om )    \nonumber   \\
& & - \sum_{\om lm}2{\rm Re}[\gamma^{\ast}_
      {\bar{\la}\s}
      \gamma_{\la \s}(\f{v_{\la}(r)}{\sq{r^2+a^2}})^2 
      \f{e^{-i(E+\om^{\rm R})T}}{(E+\om )^2}  \nonumber   \\
& & +\eta_{\bar{\la}\s}\eta^{\ast}_{\la \s}
      (\f{v^{\ast}_{\la}(r)}{\sq{r^2+a^2}})^2
      \f{e^{-i(E-\om^{\rm R})T}}{(E-\om^{\ast})^2}]
      \f{e^{\om^{\rm I}T}}{T} \}
\eeqa
for large $T \gg 1$. Here $\alpha$, $\beta$, $\gamma$, and $\eta$ are
Bogoliubov coefficients between in- and out-modes. 
This result in general shows nonvanishing 
excitations of the particle detector related to complex frequency 
modes as well as the usual contributions due to the mode mixing in real 
frequency modes. In particular, the contributions related to complex 
frequency modes are not stationary, but exponentially increasing 
in time $T$ \cite{footnote5}. The $\delta$-function dependence 
in the first two terms implies the energy conservation; that is, 
only the real frequency mode, whose quantum energy is the same as 
that of the particle detector $(\om =E)$, can excite the detector. 
For complex frequency modes, however, all modes contribute to 
the excitation possibly because the energy spectrum for any complex 
frequency mode is continuous. From the relations among Bogoliubov 
coefficients (see Ref.~\cite{Kang2}), 
one can see that it is impossible for 
$\gamma^{\ast}_{\bar{\la}\s}\gamma_{\la \s}$ and 
$\eta_{\bar{\la}\s}\eta^{\ast}_{\la \s}$ to vanish for all $\la$ 
and $\s$. Therefore, the result obtained above strongly implies
that a rotating star with ergoregion but without horizon has the quantum 
instability as well, leading exponentially time-dependent 
spontaneous energy radiation to spatial infinity. 
This quantum instability is possible because 
negative energy could be accumulated within the ergoregion.
Accordingly, our result resolves the contradiction between conclusions 
in Ref.~\cite{MDO} and Ref.~\cite{AM2}.

\vskip .5cm 
It will be straightforward to extend our formalism to other matter fields 
such as massive charged scalar and electromagnetic fields. For spinor 
fields, however, it is unclear at the present whether or not rotating 
stars with ergoregions spontaneously radiate fermionic energy to
infinity as well. It is because spinor fields do not give superradiance
in the presence of an ergoregion. 
In fact, we find that the inner product defined in Ref.~\cite{Unruh} 
for spinor fields is still positive definite in our spacetime model. 
Then, as explained below Eq.~(\ref{OrthoG}), there exists no complex 
frequency mode and hence no unstable mode for spinor fields classically. 
However, as rotating black holes give fermion emissions in the quantum 
theory inspite of no superradiance at the classical level, there might 
be some quantum process through which the ergoregion gives fermionic
spontaneous energy radiation. In addition, the main result obtained 
in the algebraic approach \cite{AM2,AM1} does not seem to depend on 
which matter field is considered. 

As shown in preceding sections, the Hamiltonian operators associated 
with unstable modes do not admit a Fock-like representation, or a vacuum 
state, or the particle interpretation of mode operators. Accordingly, 
the conventional analysis of the vacuum instability based on the uses 
of asymptotic vacua and appropriately defined number operators no longer 
applies to our case. However, we have shown that Unruh's ``particle" 
detector model, which indeed does not require the particle interpretation 
of the field, is still applicable for extracting some useful physics 
in our case. In fact, our case serves as a good example illustrating 
the point of view that the fundamental object in quantum field theory
is the field operator itself, not the ``particles" defined in a preferred 
Fock space \cite{Field}. The expectation value of the energy-momentum
tensor operator, which is defined by field operators only, should also be 
a useful quantity in our case. To obtain meaningful expectation value, 
however, renormalization of the energy-momentum tensor would have to be 
understood first in the presence of such instability
modes \cite{footnote9}. As far as we know, this interesting issue has 
never been addressed in the literature. 

\begin{figure}[hbtp]
\epsfysize=4.5cm
\hspace{5.5cm}
\epsfig{file=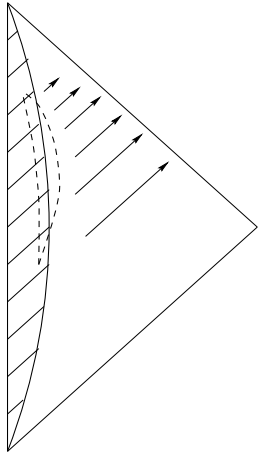,width=4cm,height=6cm}
\caption{Back reaction effect due to spontaneous evaporation of the
         ergosphere.}
\label{Ergofig2}
\end{figure}

Based on the analysis in our paper, exponentially time-dependent 
spontaneous energy radiation will occur as soon as an ergoregion is formed. 
Then the back reaction of the quantum field on the metric will change
the gravitational fields of the evolving rotational object itself, 
depending on the strength and the time scale of the spontaneous radiation.
Since the wave trapped within the ergoregion carries negative energy and 
the angular momentum in the opposite sense of the rotation, the rotating
object will loose its angular momentum and so the ergoregion can disappear
at some point of its evolution. Then the spontaneous radiation will also 
stop occurring. The corresponding Penrose diagram is shown 
in Fig.~\ref{Ergofig2}. Here the curved solid line is the trajectory of 
the surface of a rotating object. The dotted lines denote the boundaries 
of the ergoregion.

It will also be very interesting to see how our quantization procedure in
this paper can be translated into algebraic approach.
The mode decomposition constructed 
in our canonical quantization procedure suggests
that there may exist some way to construct a corresponding complex
structure in the algebraic approach. 
Finally, it should be pointed out that there are many other 
physical systems such as plasma, laser cavity, and gravitationally 
colliding system producing quasinormal modes where 
complex frequency modes play important roles. 
Generically, if a system stores some ``free"   
energy which can be released through interactions, then some amplifications
occur, revealing complex frequency modes classically. 
Therefore, the quantization formalism described in the present work may be 
useful to understand such systems in the context of quantum field theory.



\begin{thebibliography}{99}
\bibitem
{ZAMO} ZAMO's are locally nonrotating observers whose trajectories 
       are tangent to $\pa_{\tau}=\pa_t+\Om \pa_{\varphi}$ where 
       $\Om =-\pa_t \cdot \pa_{\varphi}/{\pa_{\varphi}\cdot \pa_{\varphi}}$
       and $\pa_{\varphi}$ is the rotational Killing vector field of 
       stationary axisymmetric spacetimes. The trajectories of Killing
       observers are tangent to $\pa_t$ which becomes spacelike inside
       ergoregions.
\bibitem
{footnote5} Note, however, that it does not depend on the setting time
      $t_0$ of the detector.
\end{thebibliography}
\end{document}